\RequirePackage{fix-cm}
\documentclass{svjour3}                     
\smartqed  
\usepackage{amsmath}
\usepackage{graphicx}
\usepackage{amsfonts}
\usepackage{amscd}
\usepackage[english]{babel}
\usepackage[latin1]{inputenc}
\usepackage{times}
\usepackage[T1]{fontenc}

\newcommand{\be}{\begin{equation}}
\newcommand{\ee}{\end{equation}}

\newcommand{\ba} {\begin {array}}
\newcommand {\ea} {\end {array}}
\newcommand{\bea}{\begin{eqnarray}}
\newcommand{\eea}{\end{eqnarray}}

\begin{document}

\title{Higher Order Quantum Superintegrability: a new "Painlev\'e conjecture"
}
\subtitle{Higher Order Quantum Superintegrability}


\author{Ian Marquette         \and
        Pavel Winternitz 
}


\institute{I. Marquette \at
              School of Mathematics and Physics, The University of Queensland, Brisbane, St-Lucia,
							Australia, QLD 4072 \\
              Tel.: +61 7 334 67952\\
              \email{i.marquette@uq.edu.au}           
           \and
           P. Winternitz \at
					    Centre de recherches math\'ematiques et
              \\D\'{e}partement de Math\'{e}matiques et de Statistique,
             Universit\'{e} de Montr\'{e}al,\\ CP 6128, Succ. Centre-Ville, Montr\'{e}al,
             Quebec H3C 3J7, Canada \\
              Tel.: +514-343-7271\\
              \email{wintern@CRM.UMontreal.CA }           
}


\maketitle

\begin{abstract}
We review recent results on superintegrable quantum systems in a two-dimensional Euclidean space
with the following properties. They are integrable because they allow the
separation of variables in Cartesian coordinates and hence allow a
specific integral of motion that is a second order polynomial in the
momenta. Moreover, they are superintegrable because they allow an
additional integral of order $N>2$. Two types of such superintegrable
potentials exist. The first type consists of "standard potentials" that
satisfy linear differential equations. The second type consists of "exotic
potentials" that satisfy nonlinear equations. For $N= 3$, 4 and 5 these
equations have the Painlev\'e property. We conjecture that this is true for
all $N\geq3$. The two integrals X and Y commute with the Hamiltonian, but not
with each other. Together they generate a polynomial algebra (for any $N$)
of integrals of motion. We show how this algebra can be used to calculate
the energy spectrum and the wave functions.
\end{abstract}

\section{Introduction}
\label{intro}

Let us first consider a classical system in an n-dimensional Riemannian space with Hamiltonian
\begin{eqnarray}
H=\sum_{\substack{i,k=1\\ j,k\geq 0}}^n g_{ik}(\vec{x})p_ip_k+V(\vec{x}) \,, \vec{x} \in \mathbb{R}^n.
\label{H1}
\end{eqnarray}
The system is called \emph{integrable} (or Liouville integrable) if it allows $n-1$ Poisson commuting integrals of motion (in addition to H)
\begin{eqnarray}
&X_a=f_a(\vec{x},\vec{p}) \,, \qquad a=1,\ldots,  n-1 \nonumber, \\
&\dfrac{dX_a}{dt}=\{H,X_a\}_{p}=0 \,, \{X_a,X_b\}_{p}=0.
\label{Xintegrals}
\end{eqnarray}
where $\{ ,\}_{p}$ is the Poisson bracket, and $p_{i}$ are the momenta canonically conjugate to the coordinates $x_{i}$.
\newline
This system is \emph{superintegrable} if it allows further integrals
\begin{eqnarray}
&Y_b=f_b(\vec{x},\vec{p}) \,, \qquad b=1,\ldots ,k \qquad 1\leq k \leq n-1, \nonumber \\
&\dfrac{dY_b}{dt}=\{H,Y_b\}_{p}=0 \,.
\label{Yintegrals}
\end{eqnarray}
\newline
\newline
In addition, the integrals must satisfy the following requirements :
\newline
\newline
1. The integrals $H,X_a,Y_b$ are well defined functions on phase space, i.e. polynomials or convergent power series on phase space (or an open submanifold of phase space). 
\newline
2. The integrals $H,X_a$ are in involution, i.e. Poisson commute as indicated in (\ref{Xintegrals}). The integrals $Y_b$ Poisson commute with $H$ but not necessarily with each other, nor with $X_a$.\\
\newline
3. The entire set of integrals is functionally independent, i.e., the Jacobian matrix satisfies
\begin{eqnarray}
\text{rank} \dfrac{\partial(H,X_1,\ldots,X_{n-1},Y_1,\ldots,Y_k)}{\partial(x_1,\ldots,x_n,p_1,\ldots,p_n)} =n+k
\end{eqnarray}

In quantum mechanics we define integrability and superintegrability in the same way, however in this case, $H,X_a$ and $Y_b$ are operators. The condition on the integrals of motion must also be modified e.g. as follows : 
\newline
\newline
1. $H,X_a$ and $Y_b$ are well defined Hermitian operators in the enveloping algebra of the Heisenberg algebra $ H_n \sim \{  \vec{x},\vec{p},\hbar \} $ or some generalization thereof.
\newline
2. The integrals satisfy the Lie bracket relations
\begin{eqnarray}
&[H,X_a]=[H,Y_b]=0 \,, [X_i,X_k]=0
\end{eqnarray}
\newline
3. No polynomial in the operators $H,X_a,Y_b$ formed entirely using Lie anticommutators (i.e. Jordan polynomials) should vanish identically.
\newline
\newline
The two best known superintegrable systems are the Kepler-Coulomb system with potential $V(r)=\frac{\alpha}{r}$ and the isotropic harmonic oscillator $V(r)= \alpha r^2$ \cite{p26,f35,b36,ms96}. In both cases the integrals $X_a$ correspond to angular momentum, the additional integrals $Y_a$ to the Laplace-Runge-Lenz vector for $V(r)=\frac{\alpha}{r}$ and to the quadrapole tensor $T_{ik}=p_ip_k+\alpha x_ix_k$, for $V(r)=\beta r^{2}$. No further ones were discovered until a 1940 paper by Jauch and Hill \cite{jau40} on the rational anisotropic harmonic oscillator $V(\vec{x})=\alpha \sum_{i=1}^n n_i x_i^2 \,\,, n_i \in \mathbb{Z} $. A systematic search for superintegrable systems was started in 1965 \cite{fmsuw65,w66,msvw67} and a real proliferation of them was observed during the last few years \cite{msw13}. This research program remains very active \cite{hv84,pw11,ppw12,n12,n13,ckno13,r13,msw15,behlrr16,chr17,kmp13,ekm17,lmz18,hmz17,bglv17,h18,i18}. The search has also been extended to systems with spin, magnetic fields and monopoles. Many families of superintegrable systems have been constructed using combinations of approaches such as the co-algebra \cite{bbhmr09,br98} and the recurrence method \cite{bfhkn16,msw13}. Let us just list some of the reasons why superintegrable systems are interesting both in classical and quantum physics. 
\newline  
\newline
1.In classical mechanics, superintegrability restricts trajectories to an $n-k$ dimensional subspace of phase space \cite{n72}. For $k=n-1$ (maximal superintegrability), this implies that all finite trajectories are closed and motion is periodic. 
\newline
2. Moreover, at least in principle, the trajectories can be calculated without any calculus. 
\newline
3. Bertrand's theorem states that the only spherically symmetric potentials $V(r)$ for which all bounded trajectories are closed are $\dfrac{\alpha}{r}$ and $\alpha r^2$ \cite{b73,g01}, hence no other maximally superintegrable systems are spherically symmetric. 
\newline
 4. The algebra of integrals of motion $\{H,X_a,Y_b\}$ is a non-Abelian and interesting one. Usually it is a finitely generated polynomial algebra, only exceptionally a finite dimensional Lie algebra. In the special case of quadratic superintegrability (all integrals of motion are at most quadratic polynomials in the moment), integrability is related to separation of variables in the Hamilton-Jacobi equation.
\newline
\newline
 In quantum mechanics,
\newline
\newline
1. Superintegrability leads to an additional degeneracy of energy levels, sometimes called "accidental degeneracy". The term was coined by Fock \cite{f35} and used by Moshinsky and collaborators \cite{ms96}, though the point of their studies was to show that this degeneracy is certainly no accident. Quadratic integrability is related to separation of variables to the corresponding Schrodinger equation.
Quadratic superintegrability implies multiseparability of the
Schrodinger equation.
\newline
2. A conjecture, born out by all known examples, is that all maximally superintegrable systems are exactly solvable \cite{ttw01}. If the conjecture is true, then the energy levels can be calculated algebraically. The wave functions are polynomials (in appropriately chosen variables) multiplied by some overall gauge factor. 
\newline
3. The non-Abelian polynomial algebra of integrals of motion has been obtained for various models \cite{g92,bon93,lv95,das01,gvz11,kmp13,gi13,gvz14,ekm17,lmz18,hmz17,bglv17,h18,i18}. In many cases they correspond to higher rank polynomial algebras. They provide energy spectra and information on wave functions via Casimir operators and representation theory. Moreover, it has been demonstrated how Inonu-Wigner and more generally Bocher contractions of quadratic algebras play a role \cite{kmp13,ekm17} in connecting all quadratically superintegrable models in conformally flat spaces. Interesting relations exist between superintegrability and supersymmetry in quantum mechanics \cite{m11} and even more generally other types of operator algebras appear \cite{msw18}.
\newline
4. Relation to special function theory: multivariable orthogonal polynomials, new "nonclassical" orthogonal polynomials, Askey-Wilson classification \cite{vz11,kmp13} and exceptional orthogonal polynomials \cite{ggm14,gkm10,ptv12,mq13}
\newline
The theory of superintegrable systems has also been formulated in the context of Lie theory and generalized symmetries \cite{stw01a}. As a comment, let us mention that superintegrability has also been called non-Abelian integrability. From this point of view, infinite dimensional integrable systems (soliton systems) described e.g. by the Korteweg-de-Vries equation, the nonlinear Schr\"odinger equation, the Kadomtsev-Petviashvili equation, etc. are actually superintegrable \cite{o85,o86,ow97}. Indeed, the generalized symmetries of these equations form infinite dimensional non-Abelian algebras (the Orlov-Shulman symmetries) with infinite dimensional Abelian subalgebras of commuting flows.
There is another connection between superintegrable systems in quantum mechanics and soliton theory \cite{ac91} namely the important role of the Painlev\'e property and Painlev\'e transcendents ( of second and higher order) in both.

The paper is organized as follow. In Section 2, we present the case of second order superintegrable systems in two-dimensional Euclidean space. In Section 3, we present a summary of results for integrals of motion of order N in $E_{2}$. In the Section 4, we review the case of $N=4$ with exotic potentials and separable in Cartesian coordinates and present the connection with the Chazy class of equations. We present a summary of the classification of exotic potentials with fourth order integrals separable in cartesian coordinates in Section 5. Section 6 is devoted to a discussion of the algebraic derivation of the spectrum using a cubic algebra. In Section 7 we discuss the connection with supersymmetric quantum mechanics.

\section{Second Order Superintegrability}
\label{second}
Let us consider the Hamiltonian (\ref{H1}) in the Euclidian space $E_2$ and search for second order integrals of motion \cite{fmsuw65,w66,msw13}. We have
\begin{eqnarray}
&H=\frac{1}{2} \left( p_1^2+p_2^2 \right)_{}+ V(x_1,x_2),\qquad \displaystyle X=\sum_{j+k=0}^2 \Big\{f_{jk}(x_1,x_2),p_1^jp_2^k\Big\},
\label{defH}
\end{eqnarray}
where $j,k \in \mathbb{Z}\geq 0$ and $\{, \}$ is the anti commutator.
In the quantum case we have
\begin{eqnarray}
p_j=-i \hbar \dfrac{\partial}{\partial x_j}, \qquad L_3=x_1p_2-x_2p_1.
\label{opdef}
\end{eqnarray}
The commutativity condition $[H,X]=0$ implies that the even terms $j+k=0,2$ and odd terms $j+k=1$ in X commute with H separately. Hence we can, with no loss of generality, set $f_{10}=f_{01}=0$. Further we find that the leading (second order) term in $X$ lies in the enveloping algebra of the Euclidian algebra $e(2)$. Thus we obtain
\begin{eqnarray}
X=aL_3^2+b_1(L_3p_1+p_1L_3)+b_2(L_3p_2+p_2L_3)+c_1(p_1^2-p_2^2)\\ \nonumber
+2c_2p_1p_2+\phi (x_1,x_2)
\label{inta}
\end{eqnarray}
where $a,b_i,c_i$ are constants. The terms $c_{0}(p_{1}^{2}+p_{2}^{2})$ has been removed by linear combinations with the Hamiltonian.

The function $\phi (x_1,x_2)$ must satisfy the determining equations
\begin{eqnarray}
\phi_{x_1}&=&-2(a x_2^2+2b_1x_2+c_1) V_{x_1}+2(ax_1x_2+b_1x_1-b_2x_2-c_2)V_{x_2} \nonumber \\
\phi_{x_2}&=&-2(a x_1x_2+b_1x_1-b_2x_2-c_2) V_{x_1} + 2(-ax_1^2+2b_2x_1+c_1)V_{x_2},
\label{phieq1}
\end{eqnarray}
The compatibility condition $\phi_{x_1x_2}=\phi_{x_2x_1}$ implies
\begin{eqnarray}
&&(-ax_1x_2-b_1x_1+b_2x_2+c_2)(V_{x_1x_1}-V_{x_2x_2}) \nonumber\\
&&-(a(x_1^2+x_2^2)+2b_1x_1+2b_2x_2+2c_1)V_{x_1x_1} \nonumber \\
&&-(ax_2+b_1)V_{x_1}+3(ax_1-b_2)V_{x_2}=0.
\label{compatibilityphi}
\end{eqnarray}
Eq. (\ref{compatibilityphi}) is exactly the same equation that we would have obtained if we had required that the potential should allow the separation of variables in the Schr\"odinger equation in one of the coordinate system in which the Helmholtz equation allows separation ( $V(x_{1},x_{2})=0$ in (\ref{defH})  ). Another important observation is that (\ref{phieq1}) and (\ref{compatibilityphi}) do not involve the Planck constant. Indeed, if we consider the classical functions $H$ and $X$ in (\ref{defH}) 
and require that they Poisson commute, we arrive at exactly the same conclusions and to equations (\ref{phieq1}) and (\ref{compatibilityphi}).
Thus for quadratic integrability (and superintegrability) the potentials and integrals of motion coincide in classical and quantum mechanics (up to a possible symmetrization of the integrals). The Hamiltonian (\ref{H1}) is form invariant under Euclidian transformations, so we can classify the integrals $X$ into equivalence classes under rotations, translations and linear combinations with $H$. There are two invariants in the space of parameters $a,b_i,c_i$, namely
\begin{eqnarray}
I_1=a,\qquad I_2=(2ac_1-b_1^2+b_2^2)^2+4(a c_2-b_1b_2)^2
\end{eqnarray}
Solving (\ref{defH}) for different values of $I_1$ and $I_2$ we obtain :
\begin{eqnarray}
I_1=I_2=0 & \qquad V_C=f_1(x_1)+f_2(x_2)& \/  \nonumber \\
I_1=1,\,I_2=0& \qquad V_R=f(r)+\dfrac{1}{r^2}g(\phi) & \quad x_1 =r \cos \phi ,\,x_2 =r \sin \phi \nonumber \\
I_1=0,\,I_2=1&\qquad V_P=\dfrac{f(\xi)+g(\eta)}{\xi^2+\eta^2} & \quad x_1=\dfrac{\xi^2-\eta^2}{2} \,,x_2=\xi \eta \nonumber \\
I_1=1,\,I_2=l^2 \neq 0 & \qquad V_E=\dfrac{f(\sigma)+g(\eta)}{\cos^2\sigma- \cosh^2\rho} &
\begin{array}{rl}
&x_1=l \cosh \rho \cos \sigma  \\
&x_2=l \sinh \rho \sin \sigma \\
&0 < l < \infty
\end{array}
\label{fourtypes}
\end{eqnarray}
We see that $V_C,V_R,V_P$ and $V_E$ correspond to separation of variables in Cartesian, polar, parabolic and elliptic coordinates, respectively and that second order integrability (in $E_2$) is equivalent to the separation of variables in the Hamilton-Jacobi and the Schrodinger equation. For second order superintegrability, two integrals of the form (\ref{inta}) exist and the Hamiltonian separates in at least two coordinate systems. Four three-parameter families of superintegrable systems exist namely
\begin{eqnarray}
&V_I=\alpha(x^2+y^2)+\dfrac{\beta}{x^2}+\dfrac{\gamma}{y^2},& V_{II}=\alpha(x^2+4y^2)+\dfrac{\beta}{x^2}+\gamma y \nonumber \\
&V_{III}=\dfrac{\alpha}{r}+\dfrac{1}{r^2}(\dfrac{\beta}{\cos^2\frac{\phi}{2}}+\dfrac{\gamma}{\sin^2\frac{\phi}{2}}),& V_{IV}=\dfrac{\alpha}{r}+\dfrac{1}{\sqrt{r}}(\beta \cos \frac{\phi}{2}+\gamma \sin \frac{\phi}{2})
\label{fourform}
\end{eqnarray}
The classical trajectories, quantum energy levels and wave functions for all of these systems are known. The potentials $V_I$ and $V_{II}$ are isospectral deformations of the isotropic and an anisotropic harmonic oscillator, respectively, whereas $V_{III}$ and $V_{IV}$ are isospectral deformations of the Kepler-Coulomb potential. In n-dimensional space $E_n$, a set of n commuting second order integrals corresponds to a separable coordinate system. All of the above results on quadratic superintegrability have been generalized to arbitrary dimensions, to spaces of constant curvature and to other real and complex spaces \cite{k86,kkm18,m77,msw13}.

\section{Summary of results for integrals of motion of order N in $E_{2}$}

In quantum mechanics on two-dimensional Euclidean space $E_{2}$ the most general N-th order integral has the form

\begin{equation}
X=\frac{1}{2}\sum_{l=0}^{[\frac{N}{2}]}\sum_{j=0}^{N-2l}\{ f_{j,2l},p_{1}^{j}p_{2}^{N-j-2l}\}
\label{equaXN}
\end{equation}
where $f_{j,2l}$ are real functions of x,y and we set $f_{j,2l}=0$ for $j,l<0$ or $j>N-2l$. The brackets $\{ ,\}$ denote a symmetrization. In classical mechanics the brackets are inessential. The determining  equations following from the commutativity relation $[H,X]=0$ were obtained in \cite{sw15} for arbitrary $N \geq 2$, both in the classical and quantum cases. The equations are quite complicated but completely explicit.

A priori the Lie or Poisson commutator $[H,X]$ is a polynomial of order $N+1$ in the components of the momenta $p_{i}$. The terms of order $N+1$ are linear and do not involve the potential $V(x,y)$. All lower order terms are nonlinear since they involve products of the unknown potential and the unknown coefficients $f_{j,2l}$.

An analysis of the  highest and second to highest order determining equations provides several important results.

1. Even and odd parity terms in $X$ commute with $H$ separately, so all terms in (\ref{equaXN}) have the same parity ( this is already built into eq. (\ref{equaXN})).

2. The leading terms in $X$ are polynomials of order N in
the enveloping algebra of the Euclidean Lie algebra i.e.

\begin{equation}
X=X_{L} + l.o.t
\label{equaXL}
\end{equation}
\[ X_{L}=\frac{1}{2}\sum_{0 \leq m+n \leq N} A_{N-m-n,m,n} \{ L_{3}^{N-m-n},p_{1}^{m}p_{2}^{n}\} \]
where the coefficient $A_{N-m-n,m,n}$ are real constants. Indeed the leading terms are obtained for $l=0$ in (\ref{equaXN}) and are polynomials

\begin{equation}
f_{j0}=\sum_{n=0}^{N-j}\sum_{m=0}^{j} {{N-n-m}\choose{j-m}}
A_{N-n-m,m,n}x^{N-j-m}(-y)^{j-m}
\end{equation}

3) The set of determining equations $f_{j2}$ does involve the potential and is nonlinear. However, the equations are in general incompatible. A compatibility condition for arbitrary N is the linear PDE

\begin{equation}
\sum_{j=0}^{N-1}\partial_{x}^{N-1-j}\partial_{y}^{j}(-1)^{j}[ (j+1)f_{j+1,0}\partial_{x}V + (N-j)f_{j0}\partial_{y}V]=0
\label{complinV}
\end{equation}

This is a linear PDE for V alone, since the coefficients $f_{j0}$ are already known in terms of the constants $A_{N-m-n,m,n}$. Other compatibility condition exist, but they are nonlinear PDEs for the potential $V(x,y)$ and are less useful than (\ref{complinV}).

For $N=2$ the condition (\ref{complinV}) reduces to the condition  (\ref{compatibilityphi}) and provides the connection between second order integrability and the separation of variables.

For $N \geq 3$ eq. (\ref{complinV}) is also the starting point for all further studies. Right from the beginning we distinguish two types of integrable potentials:

(i) Standard potentials. For these the linear compatibility condition LCC (\ref{complinV}) is satisfied nontrivially. For $N=2$ all integrable potentials are standard.

(ii) Exotic potentials. These exist for $N \geq 3$ and for them the LCC is satisfied trivially i.e. all coefficients $A_{N-n-m,m,n}$ that figure in the LCC vanish identically. Surprisingly that does not imply that the integral $X$ vanishes; it does however greatly simplify.

Solving the remaining nonlinear PDEs is still a formidable task for any $N \geq 3$, specially in quantum mechanics. Instead of attempting this task we turn to a simpler problem, namely construct superintegrable systems in $E_{2}$
with two independent integrals of motion $X$ and $Y$, where
X is of first or second order and Y is of the order $N$. The integrals $X$ implies that $V(x,y)$ has one of the form given in (\ref{fourtypes}). The potential in (\ref{fourtypes}) depends on two arbitrary functions of one variable. Hence the LCC (\ref{complinV}) is no longer a PDE but reduces to one or several ODEs. The most interesting cases occur when the potential has the form $V_{C}$ and $V_{R}$ of (\ref{fourtypes}) i.e. allows separation in Cartesian \cite{gw02,g04,mw07,mw08,m09a,m09b,w09,msw17,aw18,rtw08} or polar coordinates \cite{tw10,elw17,elwy18,ttw09,ttw10,pw10,ewy18}.
Let us now turn to the example of exotic potentials allowing the separation of variables in cartesian coordinates and admitting an additionl independent integral of order $N=4$.

\section{Fourth Order Superintegrability and Exotic Potentials}

The article \cite{msw17} is part of a general program the aim of which is to derive, classify, and solve the equations of motion of superintegrable systems with integrals of motion that are polynomials of finite order N in the components of linear momentum. The search has been performed in two-dimensional Euclidean space. The study of Hamiltonians with integrals of motion of order $N=3$ was started in \cite{gw02} and a classification of Hamiltonians separable in Cartesian coordinates with an integrals of order $N=3$ was performed \cite{g04}. The obtained classical and quantum Hamiltonian systems have been studied in \cite{mw07,mw08,m09a,m09b,w09}. In \cite{msw17} the case $N=4$ was considered and all exotic potentials have been classified. The connection with the Painlev\'e property and Chazy class of equations was also highlighted. Partial results which consist in classifying all doubly exotic potentials were performed for $N=5$ \cite{aw18}. Results are known for systems with integrals of arbitrary order $N$ \cite{rtw08} and anisotropic oscillator complemented by Painlev\'e transcendents \cite{m11}. In this review we concentrate on superintegrable systems with Hamiltonians of the form
\begin{equation}\label{H}
H=\frac{1}{2}(p_1^2+p_2^2)+V(x,y),
\end{equation} 
in two dimensional Euclidean space $E_2$. In classical mechanics, $p_1$ and $p_2$ are the canonical momenta conjugate to the Cartesian coordinates $x$ and $y$. In quantum mechanics, we have $p_{i}$ and $L_{i}$ in eq.(\ref{opdef}).

The determining equations for fourth-order classical and quantum integrals of motion were derived earlier and they are a special case of $N$th order ones given in \cite{sw15}. In the quantum case, the integral is $Y^{(4)}=Y:$
\begin{align}\label{Y}
Y=\sum_{j+k+l=4} \frac{A_{jkl}}{2} \{L_3^j,p_1^k p_2^l\}+\frac{1}{2}(\{g_1(x,y),p_1^2\}\\
+\{g_2(x,y),p_1p_2\}+\{g_3(x,y),p_2^2\})+l(x,y),\nonumber
\end{align}
where $A_{jkl}$ are real constants, the brackets $\{.,.\}$ denote anti-commutators and the Hermitian operators $p_1,p_2$ and $L_3$ are given in (\ref{opdef}). The functions $g_1(x,y), g_2(x,y), g_3(x,y),$ and $l(x,y)$ are real and the operator $Y$ is self adjoint. Equation (\ref{Y}) is also valid in classical mechanics where $p_1, p_2$ are the canonical momenta conjugate to $x$ and $y$, respectively (and the symmetrization becomes irrelevant). The commutation relation $[H,Y]=0$ with $H$ in (\ref{H}) provides the determining equations
\begin{subequations}\label{det1}
\begin{align}
g_{1,x}=4f_1V_x+f_2V_y,\label{det1a}\\
g_{2,x}+g_{1,y}=3f_2V_x+2f_3V_y,\label{det1b}\\
g_{3,x}+g_{2,y}=2f_3V_x+3f_4V_y,\label{det1c}\\
g_{3,y}=f_4V_x+4f_5V_y.\label{det1d}
\end{align}
\end{subequations}
These 4 equations are linear PDEs and involve 4 unknown functions $g_1,g_2,g_3,V$. Furthermore we have the following two further equations
\begin{subequations}\label{l1}
\begin{align}
\ell_{x}=&2g_1V_x+g_{2}V_y+\frac{\hbar^2}{4}\bigg((f_2+f_4)V_{xxy}-4(f_1-f_5)V_{xyy}-(f_2+f_4)V_{yyy}\nonumber\\
&+(3f_{2,y}-f_{5,x})V_{xx}-(13f_{1,y}+f_{4,x})V_{xy}-4(f_{2,y}-f_{5,x})V_{yy}\nonumber\\
& -2(6A_{400}x^2+62A_{400}y^2+3A_{301}x-29A_{310}y+9A_{220}+3A_{202})V_x\nonumber\\
& +2(56A_{400}xy-13A_{310}x+13A_{301}y-3A_{211})V_y\bigg),\label{lx}\\
\ell_{y}=&g_{2}V_x+2g_{3}V_y+\frac{\hbar^2}{4}\bigg(-(f_2+f_4)V_{xxx}+4(f_1-f_5)V_{xxy}+(f_2+f_4)V_{xyy}\nonumber\\
&+4(f_{1,y}-f_{4,x})V_{xx}-(f_{2,y}+13f_{5,x})V_{xy}-(f_{1,y}-3f_{4,x})V_{yy}\nonumber\\
&+2(56A_{400}xy-13A_{310}x+13A_{301}y-3A_{211})V_x\nonumber\\
&-2(62A_{400}x^2+6A_{400}y^2+29A_{301}x-3A_{310}y+9A_{202}+3A_{220})V_y\bigg).\label{ly}
\end{align}
\end{subequations}
The quantities $f_i,\; i=1,2,..,5$ are polynomials in x and y. They are obtained from the highest order terms in the condition $[H,Y]=0$.\\[1mm] These 2 nonlinear PDEs for $l,g_1,g_2,g_3,V$ will give nonlinear compatibility condition. Explicitly for these polynomials we have
\begin{align}\label{fi}
&f_1=A_{400}y^4-A_{310}y^3+A_{220}y^2-A_{130}y+A_{040}\nonumber\\
&f_2=-4A_{400}xy^3-A_{301}y^3+3A_{310}xy^2+A_{211}y^2-2A_{220}xy-A_{121}y\nonumber\\
&\qquad +A_{130}x+A_{031}\nonumber\\
&f_3=6A_{400}x^2y^2+3A_{301}xy^2-3A_{310}x^2y+A_{202}y^2-2A_{211}xy+A_{220}x^2 \nonumber \\
& \qquad -A_{112}y+ + A_{121}x+A_{022}\nonumber\\
&f_4=-4A_{400}yx^3+A_{310}x^3-3A_{301}x^2y+A_{211}x^2-2A_{202}xy+A_{112}x \nonumber\\
&\qquad -A_{103}y+A_{013}\nonumber\\
&f_5=A_{400}x^4+A_{301}x^3+A_{202}x^2+A_{103}x+A_{004}.\nonumber\\
\end{align}
with 15 constants $A_{jkl}$. For a known potential the determining equations (\ref{det1}) and (\ref{l1}) form a set of 6 linear PDEs for the functions $g_1,g_2,g_3,$ and $l$. If $V$ is not known, we have a system of 6 nonlinear PDEs for $g_i,l$ and $V$. In any case the four equations (\ref{det1}) are a priori incompatible. The compatibility equation is a fourth-order linear PDE for the potential $V(x,y)$ alone, namely
\begin{equation}\label{V}
\partial_{yyy}(4f_1V_x+f_2V_y)-\partial_{xyy}(3f_2V_x+2f_3V_y)+\partial_{xxy}(2f_3V_x+3f_4V_y)
\end{equation}
\[-\partial_{xxx}(f_4V_x+4f_5V_y)=0.\]
This is a special case of the $N$th order linear compatibility equation (\ref{complinV}). We see that the equation (\ref{V}) does not contain the Planck constant and is hence the same in quantum and classical mechanics (this is true for any $N$). The difference between classical and quantum mechanics manifests itself in the two equations (\ref{l1}). They greatly simplify in the classical limit $\hbar \to 0$. Further compatibility conditions on the potential $V(x,y)$ can be derived for the systems (\ref{det1}) and (\ref{l1}), they will however be nonlinear. We will not go further into the problem of the fourth order integrability of the Hamiltonian (\ref{H}). Instead, we turn to the problem of superintegrability formulated in the Introduction.

\subsection{Potentials separable in Cartesian coordinates}
We shall now assume that the potential in the Hamiltonian (\ref{H}) has the form
\begin{equation}\label{Vs}
V(x,y)=V_1(x)+V_2(y).
\end{equation}
This is equivalent to saying that a second order integral exists which can be taken in the form
\begin{equation}\label{X2}
X=\frac{1}{2}(p_1^2-p_2^2)+V_1(x)-V_2(y).
\end{equation}
Equivalently, we have two one dimensional Hamiltonians
\begin{equation}\label{H1,H2}
H_1=\dfrac{p_1^2}{2}+V_1(x), \quad H_2=\dfrac{p_2^2}{2}+V_2(y).
\end{equation}

We are looking for a third integral of the form (\ref{Y}) satisfying the determining equations  (\ref{det1}) and (\ref{l1}). This means that we wish to find all potentials of the form (\ref{Vs}) that satisfy the linear compatibility condition (\ref{V}). Once (\ref{Vs}) is substituted, (\ref{V}) is no longer a PDE and will split into a set of ODEs which we will solve for $V_1(x)$ and $V_2(y)$.\\
The task thus is to determine and classify all potentials of the considered form that allow the existence of at least one fourth order integral of motion. As in every classification we must avoid triviality and redundancy. Since $H_1$ and $H_2$ of (\ref{H1,H2}) are integrals, we immediately obtain 3 "trivial" fourth order integrals, namely $H_1^2, H_2^2,$ and $H_1H_2.$ The fourth order integral $Y$ of equation (\ref{Y}) can be simplified by taking linear combination with polynomials in the second order integrals $H_1$ and $H_2$ of (\ref{H1,H2}):
\begin{equation}\label{trivial}
Y \to Y'=Y+a_1H_1^2+a_2H_2^2+a_3H_1H_2+b_1H_1+b_2H_2+b_0, \quad a_i, b_i \in \mathbb{R}.
\end{equation}
Using the constants $a_1,a_2$ and $a_3$ we set
\begin{align}
A_{004}= A_{040}= A_{022} = 0,
\end{align}
in the integral $Y$ we are searching for. At a later stage we will use the constants $b_0$ , $b_1$ and $b_2$ to eliminate certain terms in $g_1$, $g_2$, $g_{3}$ and $l.$\\

Substituting (\ref{Vs}) into the compatibility condition (\ref{V}), we obtain a linear condition, relating the functions $V_1(x)$ and $V_2(y)$
\begin{align}\label{V1+V2}
&(-60 A_{310}+240yA_{400})V_1'(x)+(-20A_{211}+60y A_{301}-60xA_{310}+240xyA_{400})V_1''(x)\nonumber\\
&+(-5A_{112}+10yA_{202}-10xA_{211}+30xyA_{301}-15x^2A_{310}+60x^2y A_{400})V_1^{(3)}(x)\nonumber\\
&+(-A_{013}+yA_{103}-xA_{112}+2xyA_{202}-x^2A_{211}+3x^2yA_{301}-x^3A_{310}+4x^3yA_{400})V_1^{(4)}(x)\nonumber\\
&+(-60A_{301}-2140xA_{400})V_2'(y)+(20A_{211}-60yA_{301}+60xA_{310}-240xyA_{400})V_2''(y)+\nonumber\\
&(-5A_{121}++10yA_{211}-10xA_{220}-15y^2A_{301}+30xyA_{310}-60xy^2A_{400})V_2^{(3)}(y)+\nonumber\\
&(A_{031}-yA_{121}+xA_{130}+y^2 A_{211}-2xyA_{220}-y^3A_{301}+3xy^2A_{310}\nonumber\\
&-4xy^3A_{400})V_2^{(4)}(y)=0.\nonumber\\
\end{align}

It should be stressed that this is no longer a PDE, since the unknown functions $V_1(x)$ and $V_2(y)$ both depend on one variable only.\\
We differentiate (\ref{V1+V2}) twice with respect to $x$ and thus eliminate $V_2(y)$ from the equation. The resulting equation for $V_1(x)$ splits into two linear ODEs (since the coefficients contain terms proportional to $y^0,$ and $y^1$), namely

\begin{subequations}\label{*(1&2)}
\begin{align}
&210A_{310}V_1^{(3)}(x)+42(A_{211}+3A_{310}x)V_1^{(4)}(x)+7(A_{112}+2A_{211}x\nonumber\\
&+3A_{310}x^2)V_1^{(5)}(x)+(A_{013}+A_{112}x+A_{211}x^2+A_{310}x^3)V_1^{(6)}(x)=0,\nonumber\\\label{*(1&2)a}\\
&840A_{400}V_1^{(3)}(x)+(126A_{301}+504A_{400}x)V_1^{(4)}(x)+14(A_{202}+3A_{301}x\nonumber\\
&+6A_{400}x^2)V_1^{(5)}(x)+(A_{103}+2A_{202}x+3A_{301}x^2+4A_{400}x^3)V_1^{(6)}(x)=0.\label{*(1&2)b}
\end{align}
\end{subequations}

\begin{subequations}\label{**(1&2)}
Similarly, differentiating (\ref{V1+V2}) with respect to $y$ we obtain two linear ODEs for $V_2(y),$
\begin{align}
&210A_{301}V_2^{(3)}(y)-42(A_{211}-3A_{301}y)V_2^{(4)}(y)+7(A_{121}-2A_{211}y \nonumber\\
&+3A_{301}y^2)V_2^{(5)}(y)-(A_{031}-A_{121}y+A_{211}y^2-A_{301}y^3)V_2^{(6)}(y)=0,\nonumber\\\label{**(1&2)a}\\
&840A_{400}V_2^{(3)}(y)-(126A_{310}-504A_{400}y)V_2^{(4)}(y)+14(A_{220}-3A_{310}y\nonumber\\
&+6A_{400}y^2)V_2^{(5)}(y)-(A_{130}-2A_{220}y+3A_{310}y^2-4A_{400}y^3)V_2^{(6)}(y)=0.\label{**(1&2)b}
\end{align}
\end{subequations}

The compatibility condition $\ell_{xy}=\ell_{yx}$, for (\ref{lx}) and (\ref{ly}) implies
\begin{align}\label{lxy-lyx}
&-g_2 V_1''(x)+g_2 V_2''(y)+(2g_{1y}-g_{2x}) V_1'(x)+(g_{2y}-2g_{3x})V_2'(y)+\nonumber\\
&\frac{\hbar^2}{4}\bigg((f_2+f_4)(V_1^{(4)}-V_2^{(4)})+(f_{2x}-4f_1'(y))V_1^{(3)}+(4f_5'(x)-5f_{2y}-f_{4y})V_2^{(3)}\nonumber\\
&+(3f_{2yy}+4f_{4xx}+6A_{211}-26A_{301}y+26A_{310}x-112A_{400}xy) V_1''\nonumber\\
&-(4f_{2yy}+3f_{4xx}+6A_{211}-26A_{301}y+26A_{310}x-112A_{400}xy)V_2''\nonumber\\
&+(84A_{310}-360A_{400}y)V_1'+(84A_{310}+360A_{400}y)V_2'\bigg)=0.
\end{align}
This equation, contrary to (\ref{*(1&2)}) and (\ref{**(1&2)}), is nonlinear since it still involves the unknown functions $g_1$, $g_2,$ and $g_3$, (in addition to $V_1(x)$ and $V_2(y)$).\\

\subsection{ODEs with the Painlev\'e property}
In order to study exotic potentials $V(x,y)=V_1(x)+V_2(y),$ allowing fourth order integrals of motion in quantum mechanics we must first recall some known results on Painlev\'e type equations \cite{p02,g10,in56}. Painlev\'e and Gambier showed that 50 classes of second order ODE exist that are single valued about their singular points. Six of them are "`irreducible", i.e. cannot be solved in terms of linear ODEs or elliptic functions, namely:

\begin{eqnarray}
P_{1}''(z) &=& 6P_{1}^{2}(z)+z \quad , \nonumber \\
P_{2}''(z)  &=& 2P_{2}(z)^{3}+zP_{2}(z)+\alpha \quad , \nonumber \\
P_{3}(z)''	 &=& \frac{P_{3}'(z)^{2}}{P_{3}(z)}-\frac{P_{3}'(z)}{z}+\frac{\alpha P_{3}^{2}(z)+\beta}{z}+\gamma P_{3}^{3}(z)+\frac{\delta}{P_{3}(z)} \quad , \nonumber \\
P_{4}(z)''	 &=& \frac{P_{4}^{'2}(z)}{2P_{4}(z)} + \frac{3}{2}P_{4}^{3}(z) + 4zP_{4}^{2}(z) + 2(z^{2} -
\alpha)P_{4}(z) +  \frac{\beta}{P_{4}(z)} \quad , \nonumber \\
P_{5}''(z)	 &=& (\frac{1}{2P_{5}(z)}+\frac{1}{P_{5}(z)-1})P_{5}'(z)^{2}-\frac{1}{z}P_{5}'(z)+\frac{(P_{5}(z)-1)^{2}}{z^{2}}(\frac{\alpha P_{5}^{2}(z)+\beta}{P_{5}(z)})\nonumber \\
&&+\frac{\gamma P_{5}(z)}{z}+\frac{\delta P_{5}(z)(P_{5}(z)+1)}{P_{5}(z)-1} \nonumber \\ \quad 
P_{6}''(z)   &=& \frac{1}{2}(\frac{1}{P_{6}(z)}+\frac{1}{P_{6}(z)-1}+\frac{1}{P_{6}(z)-z})P_{6}'(z)^{2}-(\frac{1}{z}+\frac{1}{z-1}+\frac{1}{P_{6}(z)-z})P_{6}'(z) \nonumber \\
&& +\frac{P_{6}(z)(P_{6}(z)-1)(P_{6}(z)-z)}{z^{2}(z-1)^{2}}(\gamma_{1}+\frac{\gamma_{2}z}{P_{6}(z)^{2}}+\frac{\gamma_{3}(z-1)}{(P_{6}(z)-1)^{2}}+\frac{\gamma_{4}z(z-1)}{(P_{6}(z)-z)^{2}}) \nonumber \\
\end{eqnarray}
\newline
\newline
An ODE has the Painlev\'e property if its general solution has no movable branch points, (i.e. branch points whose location depends on one or more constants of integration). For a review and further developments see \cite{c92,h09,c99,cm08}. Passing the test \cite{ablo78} is a necessary condition for having the Painlev\'e property. We shall need it only for equations of the form

\begin{align}\label{noeq}
W^{(n)}=F(y,W,W',W'',...,W^{(n-1)}),
\end{align}
where $F$ is polynomial in $W,W',W'',...,W^{(n-1)}$ and rational in $y$. 
\newline
\newline
The general solution must have the form of a Laurent series with a finite number of negative power terms
\begin{align}\label{laurent}
W=\Sigma_{k=0}^\infty d_k (y-y_0)^{k+p}, \; d_0 \neq 0,
\end{align} 
satisfying the requirements
\begin{enumerate}
\item The constant $p$ is a negative integer. 
\item The coefficients $d_k$ satisfy a recursion relation of the form
$$P(k)d_k=\phi_k(y_0,d_0,d_1,...,d_{k-1}),$$
where $P(k)$ is a polynomial that has $n-1$ distinct nonnegative integer zeros. The values of $k_j$ for which we have $P(k_j)=0$ are called resonances and the values of $d_k$ for $k=k_j$ are free parameters. Together with the position $y_0$ of the singularity we thus have $n$ free parameters in the general solution (\ref{laurent}) of the n-th order ODE (\ref{noeq}) .
\item A compatibility condition, also called the resonance condition:
$$\phi_k(y_0,d_0,d_1,...,d_{k-1})=0,$$
must be satisfied identically in $y_0$ and in the values of $d_{k_j}$ for all $k_j; j=1,2,...,n-1.$
\end{enumerate}

This test is a generalization of the Frobenius method used to study fixed singularities of linear ODEs . Passing the Painlev\'e test is a necessary condition only. To make it sufficient one would have to prove that the series (\ref{laurent}) has a nonzero radius of convergence and that the $n$ free parameters can be used to satisfy arbitrary initial conditions. A more practical procedure that we shall adopt is the following. Once a nonlinear ODE passes the Painlev\'e test one can try to integrate it explicitly. 
\newline
Let us first investigate the cases that may lead to "exotic potentials", that is potentials which do not satisfy any linear differential equations. That means that either (\ref{*(1&2)}) or (\ref{**(1&2)}) (or both) must be satisfied trivially. The linear ODEs (\ref{*(1&2)}) are satisfied identically if we have
\begin{align}\label{*}
A_{400}=A_{310}=A_{301}=A_{211}=A_{202}=A_{112}=A_{103}=A_{013}=0.
\end{align}
The linear ODEs (\ref{**(1&2)}) are satisfied identically if we have 
\begin{align}\label{**}
A_{400}=A_{310}=A_{301}=A_{211}=A_{220}=A_{121}=A_{130}=A_{031}=0. 
\end{align}

If (\ref{*}) and (\ref{**}) both hold then the only fourth order integrals are the trivial ones $H_1^2, H_2^2$ and $H_1H_2.$ Their existence does not imply superintegrability, it is simply a consequence of second order integrability. In other words, no fourth order superintegrable systems, satisfying (\ref{*}) and (\ref{**}) simultaneously, exist. This means that at most one of the functions $V_1(x)$ or $V_2(y)$ can be "exotic". The other one will be a solution of a linear ODE. For third order integrals both $V_1(x)$ and $V_2(y)$ can be exotic.\\

Let us consider the case when, (\ref{**}) is valid and (\ref{*}) not. The leading-order term for the nontrivial fourth order integral has the form
\begin{align}\label{V2trivially}
Y_L= A_{202} \{L_3^2, p_{2}^2\}+ A_{112} \{L_{3}, p_{1} p_{2}^2\}+ A_{103} \{L_{3}, p_{2}^3\}+2 A_{013} p_1 p_2^3.
\end{align}
We proceed in several steps.
(i)Let us classify the integrals (\ref{V2trivially}) under translations (they leave the form of the potential (\ref{Vs})) invariant). The three classes are:
\begin{align}\label{classification}
I. &A_{202} \neq 0, A_{112}=A_{103}=0.\nonumber\\
II. &A_{202}=0,  A_{112}^2+A_{103}^2\neq 0, A_{013}=0,\nonumber\\
    &IIa. A_{103} \neq 0,\nonumber\\
    &IIb. A_{103}=0, A_{112} \neq 0.\nonumber\\
III. &A_{202}=A_{112}=A_{103}=0, A_{013} \neq 0.\nonumber\\
\end{align}
(ii) Let us solve the linear ODE for $V(x)$
\newline
\newline
The functions $f_i$ in (\ref{fi}) reduce to
\begin{align}\label{f1-5}
&f_{1}=f_{2}=0,\nonumber\\
&f_{3}(y)=A_{202}y^2-A_{112}y,\nonumber\\
&f_{4}(x,y)=-2A_{202}xy+A_{112}x-A_{103}y+A_{013},\nonumber\\
&f_{5}(x)=A_{202}x^2+A_{103}x.
\end{align}

we obtain two equations for $V_1(x)$ namely
\begin{subequations}\label{V1*}
\begin{align}
5A_{112}V_{1}^{(3)}(x)+(A_{013}+A_{112}x)V_{1}^{(4)}(x)=0,\label{V1*a}\\
10 A_{202}V_{1}^{(3)}(x)+(A_{103}+2A_{202}x)V_{1}^{(4)}(x)=0.\label{V1*b}
\end{align}
\end{subequations}
(They replace equations (\ref{*(1&2)})). These two equations imply $V_1^{(3)}=V_1^{(4)}=0$ unless we have 
\begin{align}\label{comp}
A_{112}A_{103}-2 A_{202}A_{013}=0.
\end{align}
The result is that $V_{1}(x)$ can have one of the following forms: $V_1(x)=0, \; ax, \; a x^2,  \; \frac{a}{x^2}+b x +c x^2$ (where $bc=0$)\\[2mm]

(iii) Let us solve the nonlinear ODEs for $V_{2}(y)$ . We first introduce an auxiliary function: 
 $W(y)=\int V_2(y) \mathrm{d} y$, \\[2mm] $\widetilde W \Leftrightarrow W+\alpha y + \beta$

Case I.  $A_{202} \neq 0, A_{112}=0; Y_L=A_{202}\{L_3^2,p_2^2\}.$\\
Let $A_{202}=1.$
We obtain
\begin{align}\label{VbII4}
&\frac{1}{2}\hbar ^2 yW^{(4)}+2\hbar^2W^{(3)}-6yW'W''-2WW''+\frac{8}{3}c_2y^3 W''-8W'^2+16c_2y^2W'\nonumber\\
&+16c_2yW-\frac{16}{3}c_2^2y^4+k_1=0,
\end{align}
integrating once we get
\begin{align}\label{VbII3}
&\hbar ^2y ^2 W^{(3)}+2\hbar^2 yW''-6y^2W'^2-4yWW'+(\frac{16}{3}c_2y^4-2\hbar ^2)W'+2W^2+\frac{32}{3}c_2y^3 W\nonumber\\
&-\frac{16}{9}c_2^2y^6+k_1y^2+k_2=0.\nonumber\\
\end{align}
The equation (\ref{VbII3}) passes the Painlev\'e test. Substituting the Laurent series (\ref{laurent}) into (\ref{VbII3}), we find $p=-1$. The resonances are $r=1,$ and $r=6,$ and we obtain  $d_0=-\hbar ^2$. The constants $d_1$ and $d_6$ are arbitrary, as they should be. We now proceed to integrate (\ref{VbII3}). Using the results of Chazy, Bureau, Cosgrove and Scoufis \cite{c11,b64a,b64b,cs93,c06,c00a,c00b}
\newline
\newline
By the following transformation
$$Y=y^2,\; U(Y)=-\frac{y}{2\hbar ^2}W(y)+\frac{c_2}{6 \hbar ^2} y^4+\frac{1}{16},$$
we transform (\ref{VbII3}) to
\begin{align}\label{chazyI}
Y^2U^{(3)}=-2(U'(3YU'-2U)-\frac{c_2}{\hbar ^2} Y (Y U'-U)+k_3Y+k_4)-YU'',
\end{align}
where
$k_3=\frac{-2 k_1-12 c_2 \hbar ^2}{64 \hbar ^4},\;k_4=\frac{- k_2}{32 \hbar ^4}.$
The equation (\ref{chazyI}) is a special case of the Chazy class I equation  
It admits the first integral
\begin{align}\label{SDIb}
Y^2U''^2=-4(U'^2(YU'-U)-\frac{c_2}{2\hbar ^2}(YU'-U)^2+k_3(YU'-U)+k_4U'+k_5),
\end{align}
where $k_5$ is the integration constant. The equation of the canonical form SD-I.b.
\newline
\newline
When $c_2$ and $k_3$ are both nonzero the solution is
\begin{align}
U=&\frac{1}{4}(\frac{1}{P_{5}}(\frac{YP_5'}{P_{5}-1}-P_{5})^2-(1-\sqrt{2\alpha})^2(P_{5}-1)-2\beta \frac{P_{5}-1}{P_{5}}
+\gamma Y \frac{P_{5}+1}{P_{5}-1}+2 \delta \frac{Y^2P_{5}}{(P_{5}-1)^2}),\nonumber\\
U'=&-\frac{Y}{4P_{5}(P_{5}-1)}(P_{5}'-\sqrt{2\alpha} \frac{P_{5}(P_{5}-1)}{Y})^2-\frac{\beta}{2Y}\frac{P_{5}-1}{P_{5}}
-\frac{1}{2}\delta Y \frac{P_{5}}{P_{5}-1}-\frac{1}{4}\gamma,\nonumber\\
\end{align}
where $P_5=P_5(Y); Y=y^2,$ satisfies the fifth Painlev\'e equation
$$P_5''=(\frac{1}{2P_5}+\frac{1}{P_5-1})P_5'^2-\frac{1}{Y}P_5'+\frac{(P_5-1)^2}{Y^2}(\alpha P_5+\frac{\beta}{P_5})+\gamma \frac{P_5}{Y}+\delta \frac{P_5(P_5+1)}{P_5-1},$$
with
$$c_2=-\hbar ^2\delta,\; k_3=-\frac{1}{4}(\frac{1}{4}\gamma ^2+2\beta \delta -\delta (1-\sqrt{2 \alpha})^2),\; k_4=-\frac{1}{4}(\beta \gamma+\frac{1}{2}\gamma (1-\sqrt{2\alpha})^2),$$
$$k_5=-\frac{1}{32}(\gamma ^2((1-\sqrt{2 \alpha})^2-2\beta)-\delta((1-\sqrt{2 \alpha})^2+2\beta)^2).$$
\newline
\newline
The solution for the potential up to a constant is
\begin{align}\label{VbP5}
V(x,y)=&\dfrac{c_{-2}}{x^2}-\delta \hbar ^2 (x^2+y^2)+\hbar ^2 \big( \frac{\gamma}{P_5-1}+\frac{1}{y^2}(P_5-1)(\sqrt{2\alpha }+\alpha(2P_5-1)+\frac{\beta}{P_5})\nonumber\\
&+y^2(\frac{P_5'^2}{2 P_5}+\delta P_5 )\frac{(2P_5-1)}{(P_5-1)^2}-\frac{P_5'}{P_5-1}-2\sqrt{2\alpha}P_5'\big)+\frac{3\hbar^2}{8y^2}.\nonumber\\
\end{align}

%
%
The list of exotic superintegrable quantum potentials in quantum case that admit one second order Cartesian and one fourth order integral is given below. We also give their fourth order integrals by listing the leading terms $Y_L$ and the functions $g_i(x,y); i=1,2,3;$ and $l(x,y).$ Each of the exotic potentials has a non-exotic part that comes from $V_1(x)$. By construction $V_2(y)$ is exotic, however in 4 cases a non-exotic part proportional to $y^2$ splits off from $V_2(y)$ and can be combined with an $x^2$ term in $V_1(x)$. We order the final list below in such a manner that the first two potentials are isotropic harmonic oscillators (possibly with an additional $\dfrac{1}{x^2}$ term) with an added exotic part. The next two are $2:1$ anisotropic harmonic oscillators, plus an exotic part (in $y$).\\
Based on previous experience, we expect these harmonic terms to determine the bound state spectrum. The remaining $8$ cases have either $\dfrac{a}{x^2}$ or $c_1x$ as their non-exotic terms and we expect the energy spectrum to be continuous.
\newline
\newline
These results also highlight how the study of higher order Painlev\'e equations plays a role in the classification of superintegrable systems with higher order integrals of motion. Classes of such equations of third, fourth and fifth order have been studied by Chazy, Bureau, Cosgrove and Scoufis \cite{c11,b64a,b64b,cs93,c06,c00a,c00b}. 

\section{Summary of the classification of exotic potentials with fourth order integrals separable in cartesian coordinates}

In this section we give a list of some of these exotic potentials and their fourth order integrals. There are 12 cases that are divided into three types. We present one case among each of them.
\newline
\newline
I. Isotropic harmonic oscillator: $Q_1^1:$ ($Y_L=\{L_3^2,p_2^2\}$)
\begin{align}\label{pot1}
V(x,y)=&-\delta \hbar ^2 (x^2+y^2)+\frac{a}{x^2}+\hbar ^2 \big( \frac{\gamma}{P_5-1}+\frac{1}{y^2}(P_5-1)(\sqrt{2\alpha }+\alpha(2P_5-1)+\frac{\beta}{P_5})\\
&+y^2(\frac{P_5'^2}{2 P_5}+\delta P_5)\frac{(2P_5-1)}{(P_5-1)^2}-\frac{P_5'}{P_5-1}-2\sqrt{2\alpha}P_5'\big)+\frac{3\hbar^2}{8y^2}.\nonumber\\
g_1(x,y)=&2y( y W'+W+\frac{1}{3}\hbar ^2\delta y^3),\quad \nonumber\\
g_2(x,y)=&-2x(3yW'+W+\frac{4}{3}\hbar ^2\delta y^3) \nonumber\\
l(x,y)=& \hbar^2 x^2(\frac{1}{4}y W^{(4)}+W^{(3)})-x^2( 3y W'+W )W''-\hbar ^2 y(\frac{4}{3}\delta x^2y^2+\frac{3}{2})W''\nonumber\\
&+(4(\frac{a}{x^2}-\hbar ^2\delta x^2)y^2-3 \hbar ^2)W'+4y(\frac{a}{x^2}-\hbar ^2\delta x^2)W+\frac{4a}{3x^2}\hbar ^2\delta  y^4-2 \hbar ^2\delta x^2(\frac{2}{3} \hbar ^2\delta y^4- \hbar^2)-2 \hbar ^4\delta y^2.\nonumber\\
W(y)=&\frac{-\hbar ^2}{2y}\left(\frac{1}{P_{5}}\left(\frac{YP_5'}{P_{5}-1}-P_{5}\right)^2-(1-\sqrt{2\alpha})^2(P_{5}-1)-2\beta \frac{P_{5}-1}{P_{5}}+\gamma Y \frac{P_{5}+1}{P_{5}-1}+\frac{2 \delta  Y^2P_{5}}{(P_{5}-1)^2}\right)+\frac{\hbar ^2}{8y}-\frac{\delta \hbar^2}{3}y^3, \nonumber
\end{align}
where $P_5=P_5(Y)$; $Y=y^2$. 
\newline
\newline
II. Anisotropic harmonic oscillator: \\
\newline
\newline
$Q_{2}^1:$ ($Y_L=\{L_3,p_1p_2^2\}$)
\begin{align}\label{pot2}
V(x,y)=&c_2(x^2+4y^2)+\frac{a}{x^2}-4 \sqrt[4]{2 c_2^3 \hbar^2}y P_4 +\sqrt{2c_2} \hbar( \epsilon  P_4'+ P_4^2) \\
g_1(x,y)=&-2 yW'-W+\frac{4}{3} c_2 y^3,\;g_2(x,y)=3xW'-4c_2 xy^2,\; g_{3}(x,y)=2c_2x^2y-2a\frac{y}{x^2}, \nonumber\\
l(x,y)=& -\frac{1}{8} \hbar^2 x^2 W^{(4)}+\frac{3}{2} x^2 W'W''-(2 c_2 x^2y^2-\frac{3 }{4}\hbar ^2)W''-2(2a \frac{y }{x^2}+2c_2 x^2 y)W'-2(\frac{a}{x^2}+c_2 x^2 )W\nonumber\\
&+\frac{8}{3}c_2y^3(c_2 x^2+\frac{a}{x^2})-2 c_2 \hbar^2y.\nonumber\\
W(y)=&\sqrt[4]{8c_2 \hbar ^6}\big(\frac{1}{8P_{4}}P_{4}'^2-\frac{1}{8}P_{4}^3-\frac{1}{2}Y P_{4}^2-\frac{1}{2}(Y^2-\alpha +\epsilon )P_{4}+\frac{1}{3}(\alpha -\epsilon )Y+\frac{\beta}{4P_{4}}\big)+\frac{4c_2}{3}y^3,  \nonumber
\end{align}
where $P_4=P_4(Y);Y=-\sqrt[4]{\frac{8c_2}{\hbar ^2}} y.$
\newline
\newline
III. Potentials with no confining (harmonic oscillator) term: 8 cases occur involving $P_1,P_2,P_3$ or elliptic functions. \\[2mm]
For confining potentials the potentials involve $P_4$ and $P_5$.
($P_6$ appears in the case of separation in polar coordinates.) 
\newline
\newline
$Q_{3}^1:$ ($Y_L=\{L_3^2,p_2^2\}$)
\begin{align}\label{pot3}
V(x,y)=&\frac{a}{x^2}+\frac{\hbar ^2}{2}(\sqrt{\alpha } P_3'+\frac{3}{4} \alpha  (P_3)^2+\frac{\delta }{4P_3^2}+\frac{\beta P_3}{2y}+\frac{\gamma}{2y P_3}-\frac{ P_3'}{2y P_3}+\frac{P_3'^2}{4 P_3^2}). \\
g_1(x,y)=&2 y^2W'+2 y W,\;g_2(x,y)=-6xyW'-2xW,\; g_{3}(x,y)=4x^2W'+2a\frac{y^2}{x^2}, \nonumber\\
l(x,y)=&\hbar^2 x^2(\frac{1}{4}y W^{(4)}+W^{(3)})-x^2( 3y W'+W )W''-\frac{3}{2} \hbar^2 yW''+(4\frac{a}{x^2}y^2-3 \hbar^2)W'+4\frac{a}{x^2}yW. \nonumber\\
W(y)=&-\frac{\hbar ^2}{2y}\big(\frac{1}{4}(y\frac{P_3'}{P_3}-1)^2-\frac{1}{16} \alpha y^2 P_3^2-\frac{1}{8}(\beta+2\sqrt{\alpha})y P_3+\frac{\gamma}{8P_3} y+\frac{\delta}{16 P_3^2} y^2\big)+\frac{\hbar ^2}{8y}  \nonumber
\end{align}
\newline
\newline
The potentials $Q_1^2, Q_3^6$ and $Q_3^7$ are in the list of quantum potentials obtained by Gravel \cite{g04} respectively $Q_{18},Q_{19}, Q_{21}$. Among the integrals of motion we have $\{L_3^2,p_2^2\}$ and $\{L_3,p_2^3\}$. These can not be obtained by commuting a third and a second order integral. Let us mention that the classical limit $\hbar \rightarrow 0$ can not be taken in the expressions for the potentias like (\ref{pot1}), (\ref{pot2}) and (\ref{pot3}). The limit is singular and must be taken in the original determining equations. In particular for $N=4$ in equations (\ref{l1})  (the other determining equations (\ref{det1}) and their linear compatibility
equation (\ref{V}) do not contain $\hbar$). In the potential $Q_{1}^{1}$, the isotropic harmonic oscillator term appears with the coefficient $-\delta \hbar^{2}$. In $Q_{2}^{1}$ the coefficient of the anisotropic harmonic oscillator is $c_{2}$. We do not attach
any importance to this fact since both $c_{2}$ and $\delta$ are arbitrary real constants ( $\hbar^{2}$ could be absorbed into $\delta$ ). Moreover, as stated above the limit $\hbar \rightarrow 0$ is not allowed in these formulas.
\newline
\newline

For a complete list of exotic potentials of the form $V(x,y)=V_{1}(x)+V_{2}(y)$ with fourth order integrals we refer to the original article \cite{msw17}.

The results can be summed up as follows:

(i) For $N=4$ one of the two $V_{a}$( a=1,2) must be standard, i.e. satisfy a linear ODE.

We choose $V_{2}(y)$ to be exotic

(ii) The exotic part satisfies a nonlinear ODE that not only passes the Painlev\'e test but actually has the Painlev\'e property. Moreover $V_{2}(y)$ can always be expressed in terms of either elliptic functions or one of the original Painlev\'e-Gambier transcendents $P_{1}$,...,$P_{5}$. The sixth transcendent does not occur. However the sixth Painlev\'e transcendent $P_{6}$ plays a crucial role when the potential allows separation in polar coordinates instead of Cartesian ones \cite{ekm17,elw17,elwy18}.

(iii) The exotic potentials may have a nonexotic part that makes them confining. For $N=4$ this occurs in one of 3 versions

\[ V(x,y)=a(x^{2}+y^{2})+\frac{b}{x^{2}}+\frac{c}{y^{2}}+ V_{E}(y)\]
\[ V(x,y)=a(x^{2}+ 4 y^{2})+V_{E}(y)\]
\[ V(x,y)=a(x^{2}+ y^{2}) + V_{E}(y) \]

where $V_{E}$ is expressed in terms of $P_{4}$ or $P_{5}$.
The nonexotic parts in other cases are nonconfining like

\[ V=\frac{a}{x}+ V_{E}(y), V=ax + V_{E}(y) \]

with $V_{E}(y)$ expressed in terms of $P_{1}$, $P_{2}$, $P_{3}$ or an elliptic function. 
We expect the confining potentials to correspond to a bound spectrum in quantum mechanics.

\section{Example of Schr\"odinger equation with Painlev\'e potential}

Let us consider the exemple of an exotic potential expressed in terms of $P_{4}$ \\
The Hamiltonian and two integrals of motion in this case are \cite{m09a,m09b}
\begin{eqnarray}
H &=& \frac{1}{2}\big[p_1^2+p_2^2+\omega^2(x^2+y^2)\big]+V_{E}(x) \nonumber \\
A &=& p_1^2-p_2^2+\omega^2(x^2-y^2)+V_{E}(x) \nonumber \\
B &=& \frac{1}{2} \big\{L_3,p_1^2\big\}+\frac{1}{2}\big\{\frac{\omega^2}{2}x^2y-3xy-3yV_{E}'),p_1\big\}- \nonumber \\
&&\frac{1}{\omega^2}\big\{\frac{\hbar^2}{4}V_{E}'''+(-\omega^2 x^2-3V_{E})(\omega x + V_{E}'),p_1\big\}
\end{eqnarray}
with
\begin{eqnarray}
V_{E} &=&\epsilon\frac{\hbar\omega}{2}P_{4}^{'}(\sqrt{\frac{\omega}{\hbar}}x) +\frac{\omega\hbar}{2}P_{4}^{2}(\sqrt{\frac{\omega}{\hbar}}x)\nonumber \\
&&+\omega \sqrt{\hbar \omega}x P_{4}(\sqrt{\frac{\omega}{\hbar}}x)+\frac{\hbar\omega}{3}(-\alpha+\epsilon), \qquad \epsilon = \pm 1 \nonumber \\
P_{4}&=&P_{4}(\sqrt{\frac{\omega}{\hbar}}x, \alpha,\beta)
\label{Vp4}
\end{eqnarray}

The integrals of motion form a polynomial (cubic) algebra, satisfying
\begin{eqnarray}
&&[A,B]=C \qquad [A,C]=16 \omega^2\hbar^2B \nonumber \\
&&[B,C]=-2\hbar^{2}A^{3}-6\hbar^{2}HA^{2}+8\hbar^{2}H^{3} 
\nonumber \\
&& \qquad + \frac{\omega^{2}\hbar^{4}}{3}(4\alpha^{2}-20-6\beta-8\epsilon\alpha)A-8\omega^{2}\hbar^{4}H \nonumber \\
&& \qquad +\frac{\hbar^{5}\omega^{3}}{27}(-8\alpha^{3}-24\alpha-36\alpha\beta+24\epsilon\alpha^{2}+8\epsilon+36\epsilon\beta)
\label{ABCalgebra}
\end{eqnarray}

\begin{eqnarray}
&&K=-16\hbar^{2}H^{4}+\frac{4\hbar^{4}\omega^{2}}{3}(4\alpha^{2}-8\alpha+4-\alpha\beta )H^{2}\nonumber \\
&& \quad -\frac{4\hbar^{5}\omega^{3}}{27}(8\alpha^{3}-24\epsilon\alpha^{2}+24\alpha+36\alpha\beta-8\epsilon-36\epsilon\beta)H \nonumber\\
&& \quad -\frac{4\hbar^{6}\omega^{4}}{3}(4\alpha-8\epsilon\alpha-8-6\beta) \quad .       
\end{eqnarray}

The algebra has a Casimir operator that is a 4th order polynomial in the Hamiltonian H (with constant coefficients). The representation theory of the algebra (\ref{ABCalgebra}) and its realization in terms of a deformed oscillator algebra is used to calculate the energy spectrum and wave functions of the system. A connection with "higher order supersymmetry" also gives the wave functions. One obtains 3 series of states with energies
\begin{eqnarray}
E_1 &=& \hbar\omega\Big(p+\frac{\epsilon+3}{3}-\frac{\alpha}{3}\Big) \nonumber \\
E_2 &=& \hbar\omega\Big(p+\frac{-\epsilon+6}{6}+\frac{\alpha}{6} +\sqrt{\frac{-\beta}{8}} \Big), \qquad \beta < 0 \nonumber \\
E_3 &=& \hbar\omega\Big(p+\frac{-\epsilon+6}{6}+\frac{\alpha}{6} -\sqrt{\frac{-\beta}{8}}\Big), \quad
\end{eqnarray}
and 3 "zero modes", all in terms of the Painlev\'e transcendent $\mathcal{P}_{IV}$. 
\newline
\newline
It has been demonstrated that this construction may not provide the appropriate number of degeneracies via algebraic approaches and these case are associated with parameters of the fourth Painlev\'e transcendents related to exceptional orthogonal polynomials. The connection has been established via generalized Hermite and Okamoto polynomials \cite{mq16}. Constructions involving other integrals and their higher order polynomial algebras have been presented elsewhere \cite{mq14}. It has been shown how more complicated patterns of finite dimensional unitary representations can provide the degeneracies in these cases \cite{mq14}.

\section{SUSYQM construction and wavefunctions}

The wave functions can be calculated using another approach that is also in essence algebraic.
Supersymmetric quantum mechanics has been studied using many approaches and the intertwining of differential operators can be traced back to Darboux and Moutard \cite{j95}.  Second order supersymmetric quantum mechanics has been introduced in \cite{ais93} and has been exploited to generate ladder operators of third order \cite{acin00,cfnn04,mn08,m09b,m11}

Let us present a construction using first and second order supersymmetry given by the following intertwining relation

\begin{equation}
H_{1}q^{\dagger}=q^{\dagger}(H_{2}+2\lambda),\quad H_{1}M^{\dagger}=M^{\dagger}H_{2}
\end{equation}

These relations correspond to a third order ladder operator

\begin{equation}
H_{1}a^{\dagger}=a^{\dagger}(H_{1}+2\lambda) \quad ,
\end{equation}
\newline
where $a^{\dagger}$ and $a$ are third order operators. 

\begin{equation}
a^{\dagger}=q^{\dagger}M, a=M^{\dagger}q
\end{equation}

similarly

\begin{equation}
H_{2}a^{\dagger}=a^{\dagger}(H_{2}+2\lambda) \quad ,
\end{equation}
\newline
where $a^{\dagger}$ and $a$ are third order operators. 
\begin{equation}
a^{\dagger}=Mq^{\dagger},\quad a=qM^{\dagger} \quad .
\end{equation}
\newline
The explicit form is the following

\begin{eqnarray}
H_{i} &=& \frac{P_{x}^{2}}{2}+V_{i}(x)\quad , \nonumber \\
  q^{\dagger}  &=& \sqrt{\frac{\hbar}{2}}\partial + W_{3}(x) \quad , \nonumber \\
	q	 &=& -\sqrt{\frac{\hbar}{2}}\partial +W_{3}(x) \quad , \nonumber \\
	M^{\dagger}	 &=& (\sqrt{\frac{\hbar}{2}}\partial + W_{1}(x))(\sqrt{\frac{\hbar}{2}}\partial + W_{2}(x)) \quad , \nonumber \\
	M	 &=& (-\sqrt{\frac{\hbar}{2}}\partial + W_{2}(x))(-\sqrt{\frac{\hbar}{2}}\partial + W_{1}(x)) \quad 
\end{eqnarray}
\newline
The potentials $V_{1}$ and $V_{2}$ correspond up to an additive constant the one given by (\ref{Vp4}) with $\epsilon=1$ and $\epsilon=-1$. Moreover, the functions $W_{1}$, $W_{2}$ and $W_{3}$ that appear in the intertwining operators ( or supercharges ) are also expressed in terms of the fourth Painlev\'e transcendent
\newline
\begin{eqnarray}
W_{1,2} &=& \sqrt{\frac{\omega}{8}}P_{4}(\sqrt{\frac{\omega}{\hbar}}x)\pm \sqrt{\frac{\hbar}{2}}P_{4}'(\sqrt{\frac{\omega}{\hbar}}x)-\frac{2\sqrt{-\beta}}{\omega}\quad , \nonumber \\
 W_{3} &=&\sqrt{\frac{\omega}{2}}P_{4}(\sqrt{\frac{\omega}{\hbar}}x)-\frac{\omega}{2\hbar}x \quad . \nonumber \\
\end{eqnarray}
\newline
The spectrum is obtained for cases when normalizable zero modes of the annihilation operator exist
\begin{equation*}
a\psi_{k}^{(0)}=0.
\end{equation*}
The energy of the zero modes are for $\epsilon=1$ associated with the three solutions of the cubic algebra
\newline
\begin{eqnarray}
\psi_{a}^{0}(x) &=& e^{\int^{\sqrt{\frac{2}{\hbar}}x}\sqrt{\frac{2}{\hbar}}W_{3}(x')dx'}\quad , \nonumber \\
\psi_{b}^{0}(x) &=&(\sqrt{\frac{2}{\hbar}}W_{2}(x)-\sqrt{\frac{2}{\hbar}}W_{3}(x))e^{-\int^{\sqrt{\frac{2}{\hbar}}x}\sqrt{\frac{2}{\hbar}}W_{2}(x')dx'}\quad , \nonumber \\
\psi_{c}^{0}(x) &=& (\frac{4\sqrt{-\beta}}{\omega}+(\sqrt{\frac{2}{\hbar}}W_{2}(x)-\sqrt{\frac{2}{\hbar}}W_{3}(x))\nonumber \\
&&(\sqrt{\frac{2}{\hbar}}W_{1}(x)+\sqrt{\frac{2}{\hbar}}W_{2}(x)))e^{-\int^{\sqrt{\frac{2}{\hbar}}x}\sqrt{\frac{2}{\hbar}}W_{1}(x')dx'}\quad .
\end{eqnarray}
\newline
with the corresponding zero modes for $\epsilon=-1$
\newline
\begin{eqnarray}
\psi_{a}^{0}(x) &=& (\frac{\omega}{\hbar}(\alpha-1)-\frac{2\sqrt{-\beta}}{\omega}+(\sqrt{\frac{2}{\hbar}}W_{1}(x)+\sqrt{\frac{2}{\hbar}}W_{2}(x)) , \nonumber \\
&&(\sqrt{\frac{2}{\hbar}}W_{1}(x)-\sqrt{\frac{2}{\hbar}}W_{3}(x)))e^{\int^{\sqrt{\frac{2}{\hbar}}x}\sqrt{\frac{2}{\hbar}}W_{3}(x')dx'}\quad ,\nonumber \\
\psi_{b}^{0}(x) &=&e^{-\int^{\sqrt{\frac{2}{\hbar}}x}\sqrt{\frac{2}{\hbar}}W_{2}(x')dx'} \quad , \nonumber \\
\psi_{c}^{0}(x) &=& (\sqrt{\frac{2}{\hbar}}W_{1}(x)+\sqrt{\frac{2}{\hbar}}W_{2}(x))e^{-\int^{\sqrt{\frac{2}{\hbar}}x}\sqrt{\frac{2}{\hbar}}W_{1}(x')dx'}\quad \nonumber \\
\end{eqnarray}
\newline
\newline
In both cases $\epsilon=1$ and $\epsilon=-1$ the complete spectrum is recovered by acting with the raising operators.
In addition the raising ladder operators also admit zero modes. However due to conflicting asymptotics we can have in total
three, two or one infinite sequence of levels. When a potential allows only one infinite sequence of energies, this potential may also possess a singlet state or doublet states
\begin{equation}
a^{+}\psi(x)=a^{-}\psi(x)=0,\quad (a^{+})^{2}\psi(x)=a^{-}\psi(x)=0
\end{equation}
\newline

\section{Conclusion}
\label{sec:concl}

This review is devoted to superintegrable quantum systems with Hamiltonians of the form (\ref{defH}) with a potential satisfying (\ref{Vs}). They allow 2 integrals of motion $\{X,Y\}$ with X  (of order 2) as in (\ref{X2}) and $Y$ (of order $N$) as in (\ref{equaXN}) and (\ref{equaXL}) ( for $N=4$ see eq.(\ref{Y}) for $N$ arbitrary see ref.\cite{sw15}). So far the cases  $N=3$, 4 and 5 have been investigated in detail \cite{msw18,gw02,g04,msw17,aw18}. Some conclusions for general N can already be drawn. The general situation can be summed up as follows.

1.	The commutator  [H,Y] is a priori a linear differential operator of order $N+1$. The coefficients of all powers must vanish simultaneously. From terms of order $N+1$ we deduce that the terms of order N in Y are contained in the enveloping algebra of the Euclidean Lie algebra e(2). Moreover, all terms in Y have the same parity (after an appropriate symmetrisation), \cite{sw15}.

2.	Terms of order  $N-1$ in  the commutator provide nonlinear determining equations for the potential $V(x,y) = V_1(x) + V_2(y)$. However, for any $N>2$ a linear compatibility condition must be satisfied. It amounts to linear ODEs for $V_1(x)$ and $V_2(y)$. These may be satisfied trivially (all coefficients equal to zero). Then we  obtain "exotic potentials". If the linear compatibility condition is satisfied nontrivially, we obtain "standard potentials". So far, for $N<7$ all standard potentials are expressed in terms of elementary functions and all exotic ones pass the Painlev\'e test \cite{ablo78}. We conjecture that this is true for all  $N$.

3.	For a different approach to superintegrable systems in $E_2$ where such systems separating in Cartesian coordinates are obtained from operator algebras in one dimension we refer to \cite{msw18}.

4.	For recent results on superintegrable  systems in $E_2$ separable in polar coordinates we refer to the original articles \cite{elw17,elwy18,ewy18}.

\begin{acknowledgements}
The research of I.\ M.\ was supported by the Australian Research Council through Discovery Early Career Researcher Award DE130101067 and Australian Research Council Discovery Project DP 160101376. The research of P.W. was partially supported by an NSERC discovery research grant.
\end{acknowledgements}


\begin{thebibliography}{}
%

\bibitem{ac91}
M.~J.~Ablowitz and P.~A.~Clarkson. Solitons, Nonlinear evolution equations and inverse scattering, Cambridge University Press (1991)

\bibitem{ablo78}
M.J.~Ablowitz, A.~Ramani, and H.~Segur. Non-linear evolution equations and ordinary differential-equations of
Painlev\'e type. {\em. Lett. al Nuovo Cimento} 23 333 (1978).

\bibitem{aw18}
I.~Abouamal, P.~Winternitz. Fifth-order superintergrable quantum system separating in Cartesian coordinates. Doubly exotic potentials, {\em J. Math. Phys.} 59 022104 (2018).

\bibitem{acin00}
A.~Andrianov, F.~Cannata, M.~Ioffe and D.~Nishnianidze. Systems with higher-order shape invariance: spectral and algebraic properties. {\em Phys.Lett.} A, 266,341-349 (2000).

\bibitem{ais93}
A.~Andrianov, M.~Ioffe and V.P.~Spiridonov. Higher-derivative supersymmetry and the Witten index, {\em Phys.Lett.} A 174, 273 (1993).

\bibitem{bbhmr09}
A.~Ballesteros, A.~Blasco, F.~J.~Herranz, F.~Musso and O.~Ragnisco, (Super)integrability from coalgebra symmetry: Formalism and applications
{\em J. of Physics: Conf. Ser.} 175 012004 (2009).

\bibitem{behlrr16}
A.~Ballesteros, A.~Enciso, F.J.~Herranz, D.~Latini, O.~Ragnisco, D.~Riglioni. The classical Darboux III oscillator: factorization, Spectrum Generating Algebra and solution to the equations of motion, {\em J. Phys.: Conf. Ser.} 670: 012031 (2016).

\bibitem{bfhkn16}
A.~Ballesteros F.J.~Herranz S.~Kuru J.~Negro. The anisotropic oscillator on curved spaces: A new exactly solvable model, {\em Annals of Physics} 373, 399 (2016).

\bibitem{br98}
A.~Ballesteros and O.~Ragnisco. A systematic construction of completely integrable Hamiltonians from coalgebras, {\em J. Phys. A: Math. Gen.} 31 3791 (1998).

\bibitem{b36}
V.~Bargmann. Zur theorie des Wasserstoffatoms, {\em Z. Phys.} 99 576 (1936)

\bibitem{b73}
J.~L.~F.~Bertrand. Th\'eoreme relatif au mouvement d'un point attir\'e vers
un centre fixe, {\em C. R. Acad. Sci.} 77 849 (1873)


\bibitem{bglv17}
H. De ~Bie, V.X.~Genest, J.-M.~Lemay and L.~Vinet, A superintegrable model with reflections on $S^{n-1}$ and the higher rank Bannai-Ito algebra. {\em J. Phys. A: Math. Theor.}  50(19) 195202 (2017).

\bibitem{bon93}
D. Bonatsos, C. Daskaloyannis and K. Kokkotas,  Quantum algebraic desription of quantum superintegrable systems in 2 dimensions. {\em Phys. Rev. A} 48(5), R23407-R3410 (1993).

\bibitem{b64a}
F.J.~Bureau. Differential equations with fixed critical points. {\em Annali di Mat. pura ed applicata}, LXIV:229-364 (1964).

\bibitem{b64b}
F.J.~Bureau. Differential equations with fixed critical points. {\em Annali di Mat. pura ed applicata}, LXVI:1-116 (1964).

\bibitem{cfnn04}
J.M.~Carballo, D.J.~Fernandez C, J.~Negro, and L.M.~Nieto, Polynomial Heisenberg algebras, {\em J. Phys.} A 37, 10349 25J (2004).

\bibitem{chr17}
J.F.~Carinena, F.J.~Herranz and M.F.~Ranada. Superintegrable systems on 3-dimensional
curved spaces: Eisenhart formalism and separability. {\em J. Math. Phys.} 58 022701 (2017).

\bibitem{ckno13}
E.~Celeghini, S.~Kuru, J.~Negro and M.A.~del Olmo. A unified approach to quantum
and classical TTW systems based on factorization {\em Ann. Phys.} 332 27-37(2013).

\bibitem{c11}
J. Chazy, Sur les \'equations diff\'erentielles du troisieme ordre et d'ordre sup\'erieur dont
l'int\'egrale g\'en\'erale a ses points critiques fixes, Acta Math. 34:317-385 (1911).

\bibitem{c99}
R.~Conte.
\newblock  The Painlev\'e Approach to nonlinear Ordinary Differential
 Equations. The Painlev\'e property, one century later, 77--180. {\em Springer, New York}, (1999).

\bibitem{cm08}
R.~Conte and M.~Musette. The Painlev\'e Handbook. {\em Springer, Berlin}, (2008).

\bibitem{c06}
C.M. Cosgrove, Higher-order Painlev\'e equation in the polynomial class II: Bureau Symbol P1, {\em Stud. Appl. Math.}, 116 321-413 (2006).

\bibitem{c00a}
C.M. Cosgrove, Higher-order Painlev\'e equation in the polynomial class I: Bureau Symbol P2, {\em Stud. Appl. Math.},  104 1-65 (2000).

\bibitem{c00b}
C.M. Cosgrove, Chazy classes {IX}--{XI} of third-order differential equations, {\em Stud. Appl. Math.}, 104 171-228 (2000).

\bibitem{cs93}
C.M.~Cosgrove and G.~Scoufis. Painlev\'e classification of a class of differential equations of the
  second order and second degree.{\em Stud. Appl. Math.}, 88 25-87 (1993).

\bibitem{das01}
C. Daskaloyannis, quadratic Poisson algebras of two-dimensional classical superintegrable systems and quadratic algebras of quantum superintegrable systems, {\it J. Math. Phys.} {\bf 42} 1100--1119 (2001)

\bibitem{ekm17}
A.M.~Escobar Ruiz, E.G.~Kalnins, W.~Miller Jr. and E.~Subag, Bocher and Abstract Contractions of 2nd Order Quadratic Algebras. {\em SIGMA} 13 013, 38 pages (2017)

\bibitem{elw17} 
A.M.~Escobar-Ruiz, J.C.~Lopez Vieyra, P.~Winternitz. Fourth order superintegrable systems separating in Polar Coordinates. I. Exotic Potentials, {\em J. Phys. A} 50(49): 495206 (2017).

\bibitem{elwy18} 
A.M.~Escobar-Ruiz, J.C.~Lopez Vieyra, P.~Winternitz and I.~Yurdusen. Fourth order superintegrable systems separating in Polar Coordinates. II. Standard Potentials, {\em J. Phys. A: Math. Theor.} 51 455202 (2018).

\bibitem{ewy18} 
A.M.~Escobar-Ruiz, P.~Winternitz, I.~Yurdusen, General Nth order superintegrable systems separating in polar coordinates, {\em J. Phys. A: Math. Theor.} 51 40LT01 (2018).

\bibitem{f35}
V.~Fock. Zur theorie des wasserstoffatoms, {\em Z. Phys.} A 98 145 (1935)

\bibitem{fmsuw65}
I.~Fris, V.~Mandrosov, J.~Smorodinsky, M.~Uhl\'{\i}\v{r}, and P~Winternitz. On higher symmetries in quantum mechanics.
 {\em Phys. Lett.}, 16 354 (1965).

\bibitem{g10}
B.~Gambier. Sur les \'equations diff\'erentielles du second ordre et du premier
  degr\'e dont l'int\'egrale g\'en\'erale est \`a points critiques fixes.
 {\em Acta Mathematica}, 33 1 1910.

\bibitem{gi13}
V.~Genest and I.~Mourad. The Dunkl oscillator in the plane: I. Superintegrability, separated wavefunctions and overlap coefficients,
{\em J. Phys. A: Math. Theor.} 46 14 145201 (2013)


\bibitem{gvz11}
V.~Genest, L.~Vinet and A.~Zhedanov, Superintegrability  in two dimensions and the Racah-Wilson algebra. 
{\em Lett. Math. Phys.} 104 931 (2011)

\bibitem{gvz14}
V.~X.~Genest, L.~Vinet and A.~Alexei, Superintegrability in Two Dimensions and the Racah-Wilson Algebra, {\em Lett. Math. Phys.} 
104 8 931 (2014)

\bibitem{g01}
H.~Goldstein, C.~P.~Poole and J.~L.~Safko. Classical Mechanics (Reading,
MA: Addison-Wesley) (2001)

\bibitem{ggm14}
D.~Gomez Ullate, Y.~Grandati, R.~Milson. Rational extensions of the quantum harmonic oscillator and exceptional Hermite polynomials,  {\em J. Phys. A: Math. Theor.} 47 015203 (2014)
	
\bibitem{gkm10}
D.~Gomez-Ullate, N.~Kamran, R.~Milson. Exceptional orthogonal polynomials and the Darboux transformation, {\em J. Phys. A} 43 (2010) 434016

\bibitem{g92}
Y.~Granovskii, I.~Lutzenko and A.Z.~Zhedanov, Mutual integrability, quadratic algebras and dynamic symmetry.
{\em Ann. of Phys.},  217 1-20, 1992

\bibitem{g04}	
S.~Gravel. Hamiltonians separable in Cartesian coordinates and third-order integrals
of motion. {\em J. Math. Phys.} 45 1003-19 (2004)

\bibitem{gw02}
S.~Gravel and P.~Winternitz. Superintegrability with third order integrals in quantum and
  classical mechanics. {\em J. Math. Phys.}, 43 5902-5912 (2002). 

\bibitem{hv84}
E.~D'Hoker and L.~Vinet, Supersymmetry of the Pauli equation in the presence of a magnetic monopole, {\em Phys. Lett. B} 137 1 72 (1984)

\bibitem{h09}
 A.~N.~W.~Hone.
\newblock Painlev{\'e} Tests, Singularity Structure and Integrability.
\newblock {\em Integrability} 245-277.
\newblock {\em Springer, Berlin Heidelberg}, (2009).

\bibitem{h18}
M.F.~Hoque, Superintegrable systems, polynomial algebra structures and exact derivations of spectra. PhD Thesis, School of Mathematics and Physics, The University of Queensland, Australia, January, 175 pages, arXiv:1802.08410 (2018).

\bibitem{hmz17}
 M.F.~Hoque, I.~Marquette and Y.-Z.~Zhang, Quadratic algebra structure in the 5D Kepler system with non-central potentials and Yang-Coulomb monopole interaction. {\em Ann. of Phys.} 380 121-134 (2017).

\bibitem{i18}
P.~Iliev, Symmetry algebra for the generic superintegrable system on the sphere. 
{\em J. High Energy Phys.} 2, 44 22 pages (2018).

\bibitem{in56}
Ince E L 1956 {\it Ordinary differential equations} Dover, New York, 574p.

\bibitem{jau40}
J.~M. Jauch and E.~L. Hill, The problem of degeneracy in quantum mechanics. Phys. Rev. 57, 641-645 (1940).

\bibitem{j95}
G.~Junker, Supersymmetric Methods in Quantum and Statistical Physics, Springer, New York, (1995).


\bibitem{k86}
E.G.  Kalnins, Separation of Variables for Riemannian Spaces of Constant Curvature, Addison-Wesley,  Reading, Massachusett (1986)  p.196

\bibitem{kkm18}
E.G Kalnins, J.M Kress and W. Miller, Separation of Variables and Superintegrability 
The symmetry of solvable systems, IOP (2018).

\bibitem{kmp13}
G.E.~Kalnins, W.~Miller Jr, S.~Post. Contractions of 2D 2nd Order Quantum Superintegrable Systems and the Askey Scheme for Hypergeometric Orthogonal Polynomials. {\em SIGMA} 9 057 28 pages (2013). 

\bibitem{c92}
M.D.~Kruskal,  and P. A. ~Clarkson.
\newblock The Painlev\'e-Kowalevski and Poly-Painlev\'e Tests for Integrability.
\newblock{\em Studies in Applied Mathematics}, 86 87-165 (1992).

\bibitem{lv95}
P.~Letourneau and L.~Vinet.  Superintegrable systems, polynomial algebras and quasi-exactly solvable Hamiltonian, {\em Ann. Phys.} 243 1 144 (1995)

\bibitem{lmz18}  
Y.~Liao, I.~Marquette and Y.-Z.~Zhang, Quantum superintegrable system with a novel chain structure of quadratic algebras. {\em J. Phys. A: Math. Theor.} 51 255201 (13pp) (2018).

\bibitem{msvw67}
A.~Makarov, J.~Smorodinsky, Kh. Valiev, and P.~Winternitz. A systematic search for non-relativistic systems with dynamical
  symmetries. {\em Nuovo Cimento A}, 52 1061-1084 (1967).

\bibitem{msw15}
A.~Marchesiello, L.~\v{S}nobl and P.~Winternitz.Three-dimensional superintegrable systems in a static electromagnetic field. {J. Phys. A} 48 395206 (2015).

\bibitem{m11}
I.~Marquette, An infinite family of superintegrable systems from higher order ladder operators and supersymmetry, {\em J.Phys.Conf.Ser.} 284 012047 (2011). 

\bibitem{m09a} 
I.~Marquette, Superintegrability with third order integrals of motion, cubic algebras, and supersymmetric quantum mechanics. I. Rational function potentials, {\em J. Math. Phys.} 50 012101 (2009).

\bibitem{m09b} 
I.~Marquette, Superintegrability with third order integrals of motion, cubic algebras, and supersymmetric quantum mechanics. II. Painlev\'e transcendent potentials, {\em J. Math. Phys.} 50 095202 (2009).

\bibitem{mq13}
I.~Marquette, C.~Quesne, New families of superintegrable systems from Hermite and Laguerre exceptional orthogonal polynomials, {\em J. Math. Phys.} 54, 042102 (2013).

\bibitem{mq16}
I.~Marquette, C.~Quesne, Connection between quantum systems involving the fourth Painleve transcendent and k-step rational extensions of the harmonic oscillator related to Hermite EOP,
Journal of Mathematical Physics 57, 052101 (2016).

\bibitem{mq14}
I.~Marquette and C.~Quesne, Combined state-adding and state-deleting approaches to type III multi-step rationally-extended potentials: applications to ladder operators and superintegrability , {\em J. Math. Phys.} 55, 112103 (2014) 

\bibitem{msw17}
I.~Marquette, M.~Sajedi, P.~Winternitz, Fourth order Superintegrable systems separating in Cartesian coordinates I. Exotic quantum potentials, {\em J. Phys. A} 50 315201 (2017).

\bibitem{msw18}
I.~Marquette, M.~Sajedi, P.~Winternitz, Two-dimensional superintegrable systems from operator algebras in one dimension, 
{\em J. Phys. A} 52, 115202 (2019).

\bibitem{mw07}
I.~Marquette and P.~Winternitz. Polynomial Poisson algebras for classical superintegrable systems
  with a third order integral of motion. {\em J.Math. Phys.}, 48 012902 1-16 (2007). (erratum 49,019907).
	
\bibitem{mw08}
I.~Marquette and P.~Winternitz. Superintegrable systems with third order integrals of motion.
 {\em J. Phys. A. Math.Theor.}, 41 303031 (2008).	

\bibitem{mn08}
Mateo and J.~Negro, Third-order differential ladder operators and supersymmetric quantum mechanics, J. Phys. A: Math. Theor. 41, 045204 (2008).


\bibitem{m77}
W.~Miller, Symmetry and Separation of Variables 
Addison-Wesley,  Reading, Massachusetts, (1977) p.285.

\bibitem{msw13} 
W.~Miller, S.~Post and P.~Winternitz. Classical and quantum superintegrability with applications. {\em J. Phys. A},  46 423001 (2013).

\bibitem{ms96}
M.~Moshinsky and Yu.~F.~Smirnov. The Harmonic Oscillator in Modern Physics (New York: Harwood Academic) (1996)

\bibitem{n72}
N.~N.~Nekhoroshev. Action-angle variables and their generalizations, {\em Trans. Moscow Math. Soc.} 26 180 (1972)

\bibitem{n12}
A.~G.~Nikitin. New exactly solvable systems with Fock symmetry, {\em J. Phys. A: Math. Theor.} 45 485204 (2012)

\bibitem{n13}
A.~G.~Nikitin. Laplace-Runge-Lenz vector for arbitrary spin, {\em	J. Math. Phys.} 54 123506 (2013)

\bibitem{o85}
Yu.~A.~Orlov and E.~I.~Shulman. Additional symmetries of the nonlinear Schrodinger equation, {\em Theor. Math. Phys.} 64 862 (1985)

\bibitem{o86}
Yu.~A.~Orlov and E.~I.~Schulman. Additional symmetries for integrable equations and conformal algebra representation, {\em Lett. Math. Phys.} 12 171 (1986)

\bibitem{ow97}
Yu.~A.~Orlov and P.~Winternitz. Algebra of pseudodifferential operators and symmetries of equations in the Kadomtsev-Petviashvili hierarchy, {\em J. Math. Phys.} 38 4644 (1997)

\bibitem{p26}
W.~Pauli. Uber das wasserstoffspektrum vom Standpunkt der neuen Quantenmechanik, {\em Z. Phys.} 36 336 (1926)

\bibitem{p02}
P.~Painlev\'e. Sur les \'equations diff\'erentielles du second ordre et d'ordre
  sup\'erieur dont l'int\'egrale g\'en\'erale est uniforme.
{\em Acta Mathematica}, 25 1-85 (1902).

\bibitem{ppw12}
I.~Popper, S.~Post and P.~Winternitz P. Third-order superintegrable systems separable in
parabolic coordinates. {\em J. Math. Phys.} 53 062105 (2012)

\bibitem{ptv12}
S.~Post, S.~Tsujimoto and L.~Vinet. Families of superintegrable Hamiltonians constructed from exceptional polynomials, 
 {\em J. Phys. A. Math.Theor.} 45 405202 (2012)

\bibitem{pw10}
S.~Post and P.~Winternitz. An infinite family of deformations of the {C}oulomb potential.
 {\em J.Phys.A.Math.Gen.}, 43 222001, (2010). 

\bibitem{pw11}
S.~Post and P.~Winternitz. A nonseparable quantum superintegrable system in 2{D} real
 {E}uclidean space. {\em J. Phys. A. Math.Theor.}
 44 162001 (2011).

\bibitem{sw15} 
S.~Post and P.~Winternitz. General Nth order integrals of motion in the {E}uclidean plane. {\em J. Phys. A},  48 405201, (2015). 

\bibitem{r13}
M.F.~Ranada. Higher order superintegrability of separable potentials with a new
approach to the Post-Winternitz system {\em J. Phys. A-Math. Theor.} 46 125206 (2013)


\bibitem{rtw08}
M.A.~Rodriguez, P.~Tempesta, and P.~Winternitz. Reduction of superintegrable systems: The anisotropic harmonic oscillator. {\em Phys. Rev. E}, 78 046608 (2008).

\bibitem{stw01a}
M.~B. Sheftel, P.~Tempesta, and P.~Winternitz. Recursion operators, higher order symmetries and superintegrability in quantum mechanics. {\em Czech J. Phys.}, 51 392-399 (2001).

\bibitem{ttw01}
P.~Tempesta, A.~V. Turbiner, and P.~Winternitz. Exact solvability of superintegrable systems.
 {\em J. Math. Phys.}, 42 4248-4257, (2001).

\bibitem{ttw09}
F.~Tremblay, A.~V. Turbiner, and P.~Winternitz. An infinite family of solvable and integrable quantum systems on a plane. {\em J.Phys.A.Math.Theor.}, 42 242001, (2009). 

\bibitem{ttw10}
F.~Tremblay, A.~V. Turbiner, and P.~Winternitz. Periodic orbits for a family of classical superintegrable systems.
 {\em J.Phys.A.Math.Theor.}, 43 015202, (2010).

\bibitem{tw10}
F.~Tremblay and P.~Winternitz. Third order superintegrable systems separating in polar coordinates.
 {\em J.Phys.A. Math.Theor.}, 43 175206, (2010).

\bibitem{vz11}
L.~Vinet and A.~Zhedanov, A "missing" family of classical orthogonal polynomials, 
{\em J.Phys.A. Math.Theor.} 44 8 085201 (2011)


\bibitem{w09}
P.~Winternitz. Superintegrability with second and third order integrals of motion.
 {\em Phys.Atom.Nuclei}, 72 875-882, (2009).

\bibitem{w66}
P.~Winternitz, J.~Smorodinsky M.~Uhli\v{r} and I.~Fri\v{s}. Symmetry groups in classical and quantum mechanics.
{\em Yad. Fiz}, 4 625-635, 1966 ( English translation {\em Sov. J. Nucl. Phys.} 4, 444-450 (1967))





\end{thebibliography}


\end{document}